\documentclass{article}

\usepackage{graphicx}

\begin{document}

\title{Multifractality in the Random Parameters Model}
\author{Camilo Rodrigues Neto\footnote{camiloneto@usp.br} and 
Andr\' e C.R. Martins\footnote{amartins@usp.br} \\
\\GRIFE -- Escola de Arte, Ci\^encias e Humanidades,\\
 Universidade de S\~ao Paulo,  Av. Arlindo Bettio 1000, \\03828-000 S\~ao
 Paulo, Brazil}

\maketitle

\begin{abstract}

The Random Parameters model was proposed to explain the structure of the covariance 
matrix in problems where most, but not all, of the eigenvalues of the covariance 
matrix can be explained by Random Matrix Theory. In this article, we explore other 
properties of the model, like the scaling of its PDF as one take larger scales. 
Special attention is given to the multifractal structure of the model time series,
which revealed a scaling structure compatible with the known stylized facts for a reasonable 
choice of the parameter values.

\end{abstract}
%\pacs{}
%\keywords{Correlation Matrix; Multifractality; Random Matrix Theory; Time Series}

\section{Introduction}

The problem of determining the correct structure of the correlation
matrix is an important one in several different applications. In order
to explain that structure of the correlations in different problems,
Random Matrix Theory (RMT)~\cite{wigner, mehta} has been used in many
areas, such as magnetic resonance images~\cite{sengupta},
meteorology~\cite{santhanam}, and financial time
series~\cite{laloux99, plerou99}. This suggests that much of the
structure of the correlation matrix is due to noise.

In Finance applications, the estimation of the correlations is a
fundamental for portfolio choice~\cite{markowitz}. However, RMT does
not claim to explain all the eigenvalue spectrum of financial time
series, since a few large eigenvalues remain outside its scope. Also,
a number of results have been observed that are not in perfect
agreement with RMT, such as the observation that noise eigenvalues
seem to be a little larger than expected~\cite{kwapien2006} and that
correlations can be measured in the supposedly random part of the
eigenvalue spectrum~\cite{burda, burdafinancial}. It has also been
verified different behaviors of the eigenvalues corresponding to
different points of time, suggesting that non-stationary effects might
play an important role~\cite{drosdz2000, drosdz2001}.

The Random Parameter model, recently  proposed by one of the 
authors~\cite{martins2007a,martins2007b}, tries to fit the complete
 structure of the correlation matrix, based on
parameters that can be interpreted as typical observations of the
system.
However, more than just explaining the covariance structure, a model for 
financial time series should also
exhibit other empirical properties called stylized facts \cite{rama2001}, 
such as  volatility clustering, fat tails and multifractal long range
correlations. Here, we will explore the consequences of the 
Random Parameter model under those aspects.
The former two (volatility clustering and fat tails) are studied with usual 
statistical approaches, 
whereas the later (multifractal correlations), can be studied using 
the autocorrelation functions, power spectral
densities (either from Fourier or wavelets transforms) and
probability distribution functions. In addition, the fractal and the
multifractal analysis provide more insights on the scaling exponents. 
Here we use the singularity spectrum obtained from the Wavelet 
Transform Modulus Maxima (WTMM) method \cite{muzy91, arneodo95} 
to determine the multifractal structure of signals generated by this 
model.

% There are several ways to characterize the long-range
% correlations from the real time series and from its models. 

The paper is organized as follows. In Section 2 we describe the
model and present some statistical properties related to the PDF's
and the moments for the simulated return time series. 
Section 3 explains the multifractal concept and the method used to 
detect it. The results of the multifractal analysis are discussed in 
the Section 4. Finally, in the last section we present our conclusions.

\section{The Model}

Here, the returns $\mu_{i}$ and the correlation matrix $P_{il}$, where
both $i=1,\cdots, N$ and $l=1,\cdots, N$ refer to the assets, will
obtained from a $N\times P$ matrix $\mathbf \Phi$, that can be a
function of the time $t$, $\mathbf \Phi$(t). The matrix $\mathbf \Phi$
components $\varphi_{ij}$, where $i=1,\cdots , N$ represents the
different assets and where each value of $j$, $j=1,\cdots , P$,
$P\geq3$, can be seen as a collection of $P$ vectors $\mathbf
\varphi$, each with $N$ components. Each one of those vectors
represents a possible, typical state of the system and they are
divided in two types of vectors, $M$ main vectors, corresponding to
the true parameters of the model, and $R$ secondary ones, randomly
drawn at each time as explained bellow, where $M+R=P$.

Given $\mathbf \Phi$ and the average return vector $\mathbf \mu$, the
covariance matrix $\mathbf \Sigma$ and the correlation matrix $\mathbf
P$ will be given by

\[
\mu_{i} = E\left[ \varphi_{i} \right]= \frac{1}{M} \sum_{j=1}^{M} \varphi_{ij}
\]
\begin{equation}\label{eq:parametrization}
\Sigma_{il} = \frac{1}{M} \sum_{j=1}^{M} \varphi_{ij}\varphi_{lj} - \mu_{i}\mu_{l},
\end{equation}
\[
P_{il}=\frac{\Sigma_{il}}{\sqrt{\Sigma_{ii}\Sigma_{ll}}}.
\]
\noindent
The observed (simulated) returns $r_{i}(t)$, at instant $t$, are generated, as usual,
by a multivariate normal $N(\mu,\Sigma)$ likelihood.

In this model, the $M$ vectors are associated with the permanent,
non-random, eigenvalues of the correlation matrix. The $R$
pseudo-parameters will cause the return vector and correlation matrix
to change in time, even in the stationary case where each of the
$\varphi_{ij}$ elements are held constant (at least, for finite values
of $R$) and are the responsible for the bulky region of the
correlation matrix eigenvalues. They are drawn according to a Normal
distribution $N(0,\Sigma_{ii})$ for each asset $i$.

An interesting aspect of this choice of parameters is that it allows
for the introduction of non-stationarity in the main paremeters, while
preserving, by definition, all the properties of the correlation
matrix. That can be done by making each of the components
$\varphi_{ij}$ follow a random walk, such as
$\varphi_{ij}(t+1)=\varphi_{ij}(t)+\sigma_\epsilon$. However, for long
periods of time, this causes the variance associated with each asset
to explode and a mean reversal term becomes necessary. 

\begin{equation}\label{eq:meanrevrandomwalk}
\varphi_{ij}(t+1)=(1-\alpha)\varphi_{ij}(t)+\sigma_\epsilon,
\end{equation}
where $\alpha$ is a small number that measures the strength of the
mean-reversal process ($\alpha=0$ corresponds to no mean-reversal, while
$\alpha=1$ means that the system has no memory of its previous states). 

The choice of $\alpha$ and $\sigma_\epsilon$ is equivalent to a choice of
a value for variance for the $\varphi_{ij}$ that tends to remain the 
same for the next time instant. This can be seen by
calculating the variance of Equation~\ref{eq:meanrevrandomwalk} and
equating the variances of $\varphi_{ij}$ for $t$ and $t+1$ and one
obtains that the variances tend to the point 

\begin{equation}\label{eq:variance}
\sigma_{\epsilon}^{2}/(2\alpha-\alpha^2).
\end{equation}
This point corresponds to the variance value around which the variance
of $\varphi_{ij}$ will oscillate.

One important particular case for the purpose of the present analysis is when we make
$\alpha=\sigma_\epsilon=R=0$. With this choice, the $M$ main vectors will
not change in time and, since no random vectors are introduced, this
choice means that the average returns and correlation matrix will also not
change and we have a simple random walk model for the observed returns,
following a Normal distribution, with a correlation matrix with $M-1$
non-zero eigenvalues. This fact can be used to make comparisons with the
cases where the returns and correlation matrix do change.

The basic statistical characterization of the actual model was presented 
in \cite{martins2007a,martins2007b}.
The next section introduces the wavelet transform modulus maxima method
for multifractal analysis.

\section{Multifractal analysis}
\label{multi}

The multifractal scaling analysis have been largely used in the study of turbulence
and financial markets \cite{frisch95,financeiro}.
A first access to the multifractal scaling can be done using the 
Structure Function (SF) analysis. 
The SF are defined in the following way \cite{sf}:
\begin{equation}
  S_q(Y,\tau) \equiv \langle |Y(t+\tau) - Y(t)|^q \rangle \propto
  \tau^{q h(q)},
 \label{scaling}
\end{equation}
where $Y_{i}(t)=\sum_{{t^\prime}=0}^t r_{i}(t^\prime)$, $\tau$ is the scale and
$\langle~\rangle$ denotes the ensemble average.
The SF can be regarded as a generalization of the correlation functions
(when $q=2$).
We will also refer to $\tau$ as the scale of analysis.
For a signal that is scale invariant and self-similar, the signal is said fractal
when $h(q)$ has the same value for all $q$, otherwise, multifractal
\cite{paladin87,feder88}.
The scaling exponent $h$ is known as the H\"older exponent and, although
can be computed using the Structure Function approach \cite{sf},
this method has the disadvantage of not being able to obtain the
the scalings of the negative moments.
Another feature of the SF methods is its capability to
identify nonstationarity in the data. For stationary time series the exponent of
$S_q(r_i,\tau)$ is zero, due to the translational invariance of all statistics.

To obtain the full multifractal spectrum, i.e. positive and negative $q$ moments,
we make use of the Wavelet Transform Modulus Maxima (WTMM). 
The wavelet family used in this paper were the
$n^{th}$-derivative of Gaussian (DOG$n$), whose wavelet transform has $n$
vanishing moments and removes polynomial trends of order $n-1$ from the
signal. Because the scaling properties of the signal are preserved by the
wavelet transform, it is possible to obtain its multifractal spectrum using
this method. The number of vanishing moments for the wavelet basis ($n$) is
chosen to match the order ($n-1$) of the polynomial trends in the signal.

The wavelet transform of a signal $Y(t)$ is defined as:
\begin{equation}
  T_{\psi}(\tau,b_0)=\frac{1}{\tau}\sum^{N}_{t=1} Y(t)\psi^*
  \Big(\frac{t-b_0}{\tau}\Big),
\end{equation}
where $\tau>0$ is the scale being analyzed, $\psi$ is the mother wavelet and
$N$ is the number of discretized time steps.  
In this paper, we used $n=4$ for all analyses.

The statistical scaling properties of the singular measures found in time
series are characterized by the singularity spectrum, $D(h)$, of
the H\"older exponents, $h$, obtained 
with the WTMM method \cite{muzy91, arneodo95}, by the following equations:
\begin{equation}
  h(q)=\lim_{\tau \rightarrow 0}\frac{1}{\ln \tau}\sum_{\{ b_i(\tau) \}}
  \hat{T}_{\psi}[q;\tau,b_i(\tau)] \ln \left |T_{\psi}[\tau,b_i(\tau)] \right |
  = \lim_{\tau \rightarrow 0}\frac{1}{\ln \tau} Z(q;\tau)
  \label{wtmm_a}
\end{equation}
\begin{equation}
 D(h)=\lim_{\tau \rightarrow 0}\frac{1}{\ln \tau}\sum_{\{ b_i(\tau) \}}
  \hat{T}_{\psi}[q;\tau,b_i(\tau)] \ln \big |\hat{T}_{\psi}[q;\tau,b_i(\tau)] \big |
  = \lim_{\tau \rightarrow 0}\frac{1}{\ln \tau} Z^*(q;\tau)
  \label{wtmm_b}
\end{equation}
where
\begin{equation}
  \hat{T}_{\psi}[q;\tau,b_i(\tau)] =  \frac{\left | T_{\psi}[\tau,b_i(\tau)] \right |^q}
  {\sum_{\{ b_i(\tau) \}} \left | T_{\psi}[\tau,b_i(\tau)] \right |^q}
\end{equation}
and the summing is over the set of the WT modulus maxima \cite{footnote}
at scale $\tau$, ${\{b_i(\tau) \}}$. 
The singularity spectrum, i.e., the dependence of $D(h)$ with the
H\"older exponents, $h$, is obtained from the scaling range on the linear-log plots of
Equations (\ref{wtmm_a}) and (\ref{wtmm_b}).
The whole procedure is now a standard \cite{superficie2006} and, for brevity, 
it is not repeated here.

One way to interpret the multifractal spectrum in a physical sense is by
comparison with the Hurst exponents expected for known signals, for
instance, the \textit{fractional Brownian motion}
\cite{feder88,mandelbrot68}. 
The fractional Brownian motion can be
classified following the probabilities of its fluctuations: the usual
Brownian motion, obtained from the integration of a Gaussian distributed
white noise, has the same probability of having positive or negative
fluctuations and has $H=0.5$. 
A fractional Brownian motion with $H < 0.5$ is more
likely to have the next fluctuation with opposite sign of the last one --
it is said to be antipersistent.
Reversely, a fractional Brownian motion
with $H > 0.5$ is more likely to have the next fluctuation with the
same sign of the last one -- it is said to be persistent. 
Antipersistent signals have more local fluctuations and seem to be more irregular in small
scales. Their variance diverges slower with time than the variance of
persistent signals. The latter ones fluctuate on larger scales and seem to be smoother. 
This discussion is done in \cite{superficie2006} and a similar,
but more detailed interpretation is given in \cite{whatcolor2000}.

\section{Simulations and Results}

In order to study the multifractality effects for different values of
$\alpha$ and $\sigma_\epsilon$, simulations of time series with $2^{15}+2^{14}$ 
observations each, where the first $2^{14}$ were ignored as transient, were performed. 
The parameters $\alpha$ and $\sigma_\epsilon$ were chosen in a way that they follow the 
curve that keeps the variance constant, that is $\sigma_\epsilon$ was calculated from $\alpha$ by
Equation~\ref{eq:variance}. 
The different values for $\alpha$ were chosen  in the interval between $\alpha=0$ (and therefore $\sigma_\epsilon=0$,
meaning the true parameters remain constant) and $\alpha=1$, when there is no memory and
the previous values of the true parameters are forgotten at each time step and new 
ones are generated. Different numbers of random vectors ($R=1$, 2, 5 and 10) were also chosen for the 
simulations, in order to investigate the effects of introducing more randomness in the
model (as $R\rightarrow\infty$, the model tends to the Random Matrix Theory model
for the problem, except for the bulk eigenvalues). 

\begin{figure}[bbb]
 \includegraphics[width=12.5cm]{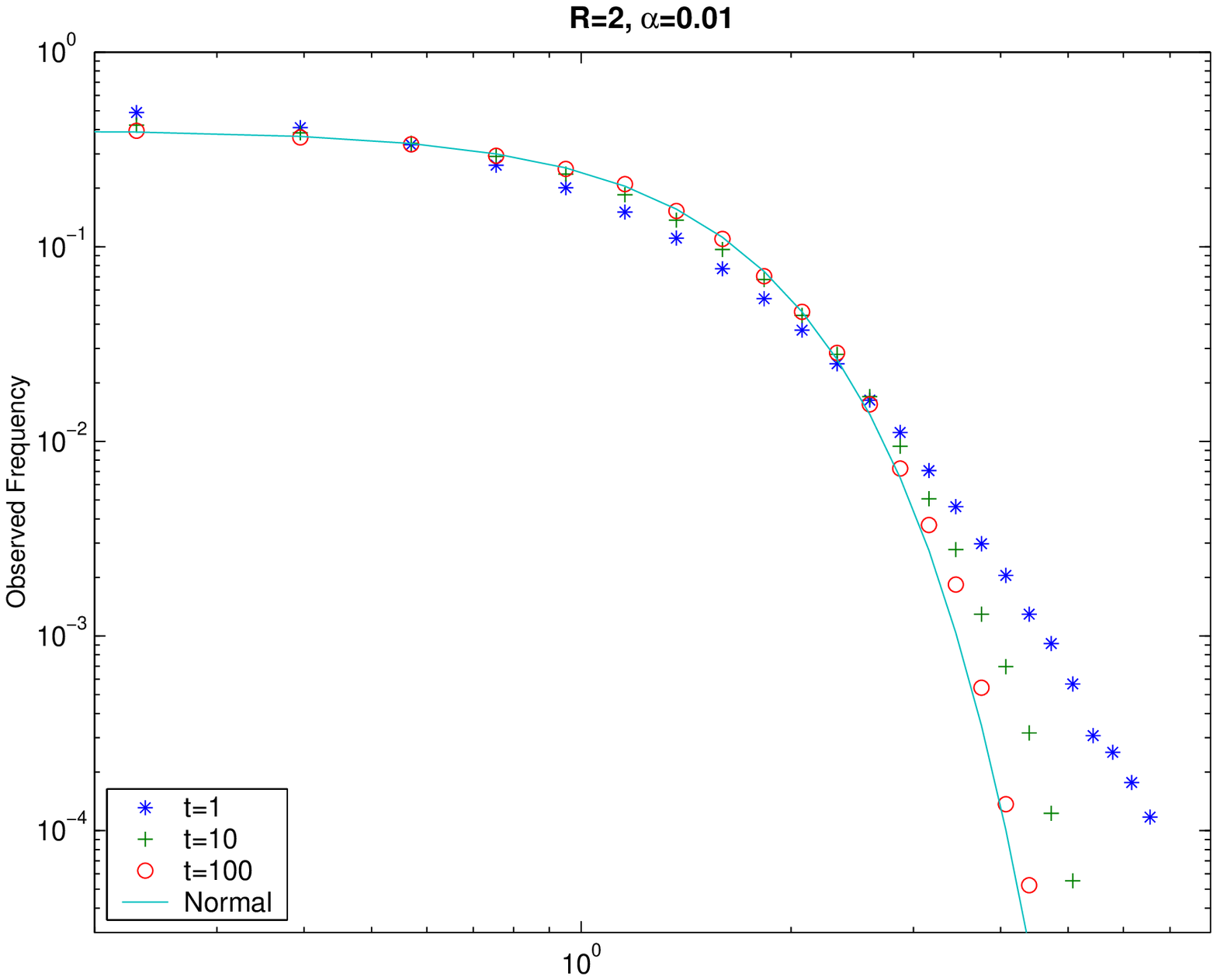}
  \caption{The tails of the observed probability distributions, averaged over 10 realizations, for different time windows ($t=1$, $t=10$ and $t=100$) for the case $R=2$ and $\alpha=0.01$.}
\label{Fig:pdfs}
\end{figure}

An interesting property of the model is how the observed distributions scale when we 
take larger time . For a window of one observation, $t=1$, the model clearly shows 
fat tails. As expected, if we take windows of larger size, the distributions converge to 
the Normal. 
Figure~\ref{Fig:pdfs} shows the observed upper tail of the distribution for the different
window sizes and it is typical of what we observe for different values of the parameters,
even though the convergence may happen at different rates. The curves correspond to the 
average over 10 different realizations of the
problem. It is easy to see that for larger windows, the curve tends to the Normal 
distribution.

\begin{figure}[bbb]
 \includegraphics[width=12.5cm]{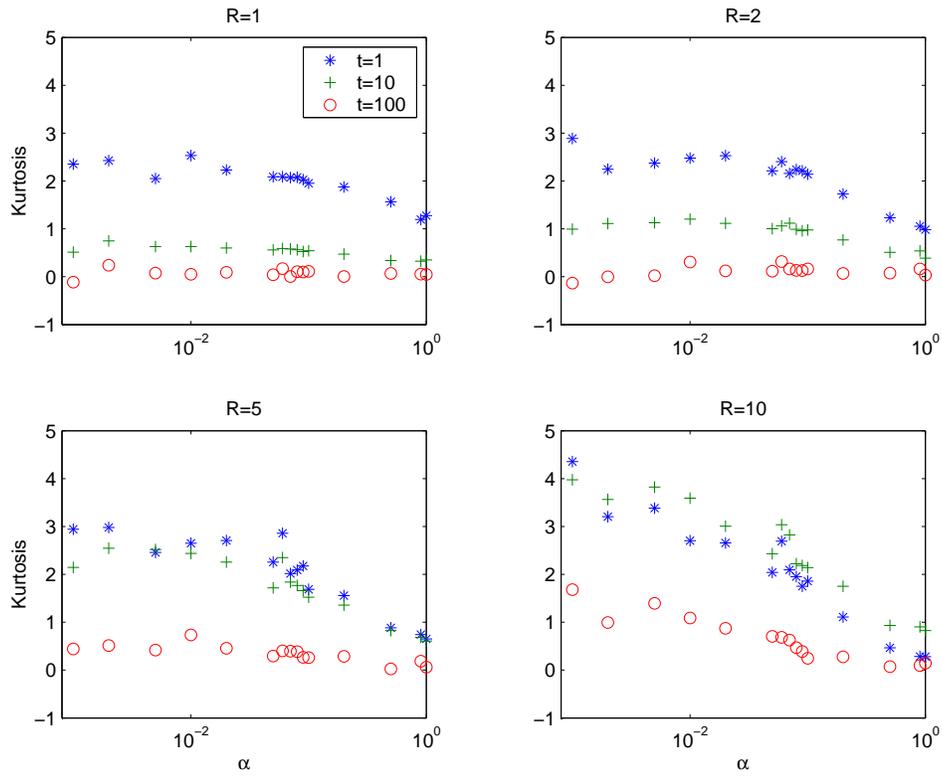}
  \caption{Kurtosis as a function of $\alpha$ for different values of $R$, averaged over 10 realizations. Curves for time windows of size $t=1$, $t=10$ and $t=100$  are shown.}
\label{Fig:kurtosis}
\end{figure}

The effect of the choice of the values of $R$ and $\alpha$ on the tails can be better
observed in Figure~\ref{Fig:kurtosis}, where average values of the kurtosis (averaged 
over the 10 realizations) are shown. Again, as the window size increases, we get closer 
to the Normal distribution, with kurtosis of 0. However, other properties can also be 
observed. Notice that, as $R$ increases, so does the kurtosis for most values of $\alpha$.
Near $\alpha=1$, the kurtosis becomes closer to zero for all cases. It is interesting to notice 
that, as was observed in a previous work~\cite{martins2007b}, when $R$ grows, the 
eigenvalues of the correlation matrix tend to those of a Random Matrix model, suggesting that
large values of $R$ might be associated with a gaussian noise. But Figure~\ref{Fig:kurtosis} 
shows it clearly that the observed noise has much fatter tails. The obvious conclusion is that,
in that region, while the covariance structure tends to the results of Random Matrix, as
$R$ grows, the PDFs have much fatter tails than a gaussian PDF.

\begin{figure}[bbb]
 \includegraphics[width=12.5cm]{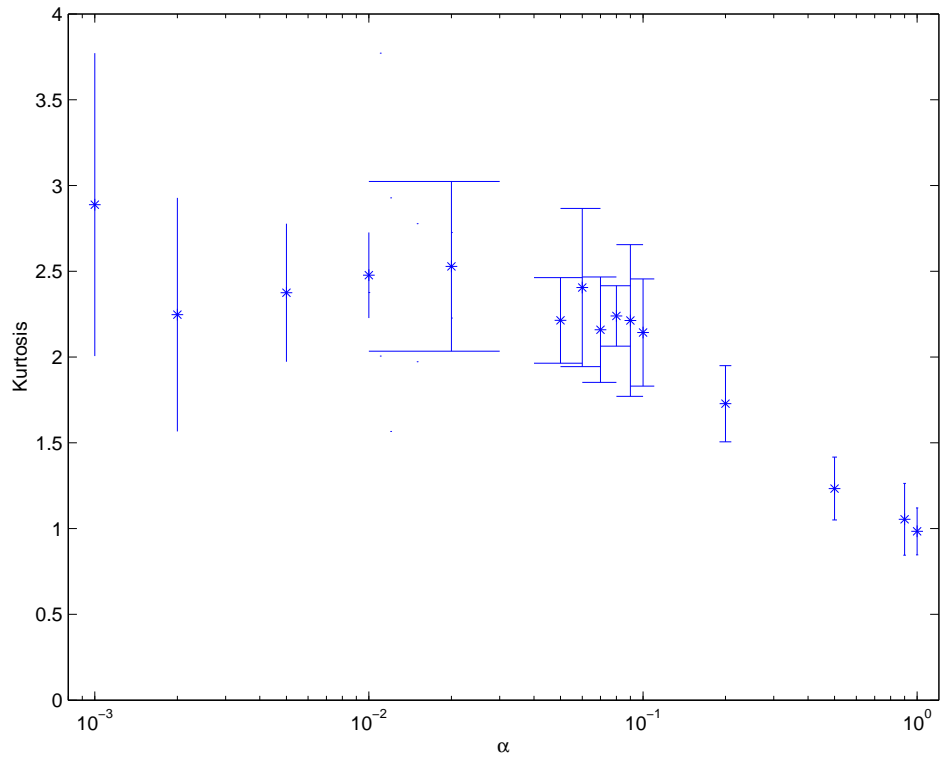}
  \caption{Kurtosis as a function of $\alpha$ for different values of $S=2$, averaged over 10 realizations, with error bars corresponding to the standard deviation of the observed kurtosis over the realizations.}
\label{Fig:kurtosiserror}
\end{figure}

The error bars associated with
the different realizations are not shown in Figure~\ref{Fig:kurtosis} to avoid cluttering.
Instead, Figure~\ref{Fig:kurtosiserror} shows the errors for the averages, when $S=2$, for
a time window of one. Notice that the error is large for small values of $\alpha$ (meaning 
that different realizations provide different values for the kurtosis) and decreases 
as  $\alpha$ goes closer to 1. This behavior is typical and observed in the other cases.

The fact that a larger value of $R$ means that larger time windows are necessary to ensure 
convergence to a Normal distribution can be explained by the fact that the random parameters 
are randomly drawn using the covariance matrix of the previous time instant. Since the $R$ random
vectors influence this matrix, as $R$ grows larger, the changes in the volatility will take a
larger time to happen and this should make the convergence slower, as it was observed.
This fact suggests that large values of $R$ are useful for describing higher frequency data, 
while smaller values of $R$ should fit lower frequency observations better. 

Another interesting step in the exploration of the model 
was to check the stationarity of the returns time series, $r$.
One signal is said stationary if it is statistically invariant under translations,
what can be evaluated using the SF approach \cite{sf}.
A stationary signal will have a flat, horizontal, structure function plot, $S_q \times \tau$.
As it is evident from Eq.~\ref{scaling}, a stationary signal presents $h = 0$ for all values 
of $q$ and $\tau$ ranges. 
This procedure determines the range of stationary scales for $\tau$, 
inside which we should look
for the scaling regimes of the price time series, $Y$. 
If there is such a scale range, we then check for the linear or non-linear behavior  
of $q.h(q)\times q$, what reveals, respectively, the fractal or multifractal 
time series dynamics.
For all values of $R$ studied, the model presented multifractal scaling for
intermediary values of $\alpha$.
Although the SF approach had shown the multifractal character of the model,
it is not able to work with negative values of $q$, necessary to obtain the full 
multifractal spectrum, what was done with the WTMM technique, described  
previously, in section \ref{multi}. 

\begin{figure}[bbb]
 \includegraphics[width=12.5cm]{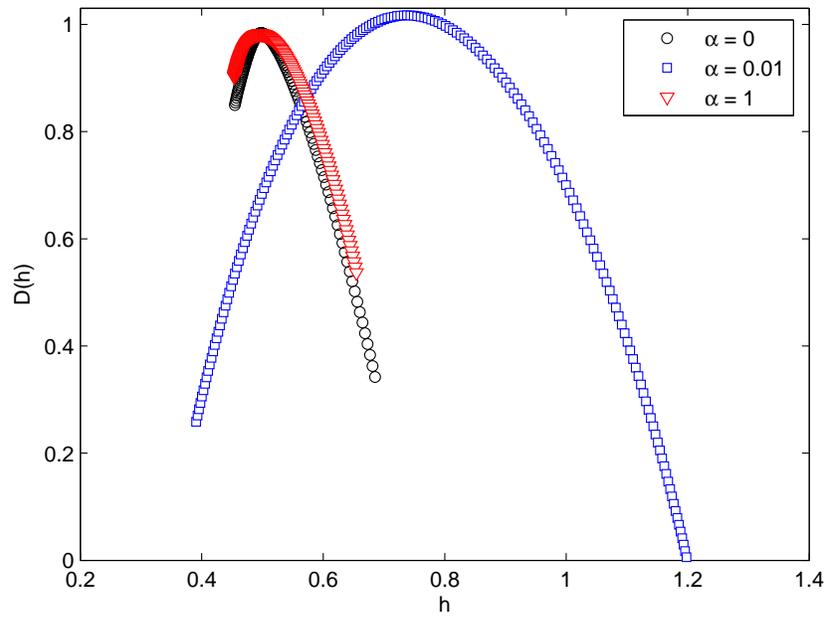}
  \caption{Multifractal spectrum for the model ($R=1$) with $\alpha=0$, $\alpha=0.01$ and
 $\alpha=1$.
It worths to note that the multifractal spectrum for $\alpha=0$ is the same of a 
fractal Brownian motion with $H=0.5$.
All the analyzed signals had 32k points long.}
\label{mfs}
\end{figure}

To proceed with the analyses, we used the WWTM method to obtain
the multifractal spectrum of $Y$, presented in Fig.~\ref{mfs}.
This figure shows the multifractal spectrum for $\alpha = 0,\: 0.01$ and $1$.
For both extrema, the MS spectra are much narrower than for the intermediary value
$\alpha = 1$.
Both techniques, SF and WTMM, had been successfully applied in the study of simulated 
and experimental time series \cite{superficie2006,chuva1999} and the
intermediary steps to obtain the multifractal spectrum will not be shown 
in detail here. The interested reader is referred to the cited bibliography.

\begin{figure}[bbb]
  \includegraphics[width=6.5cm]{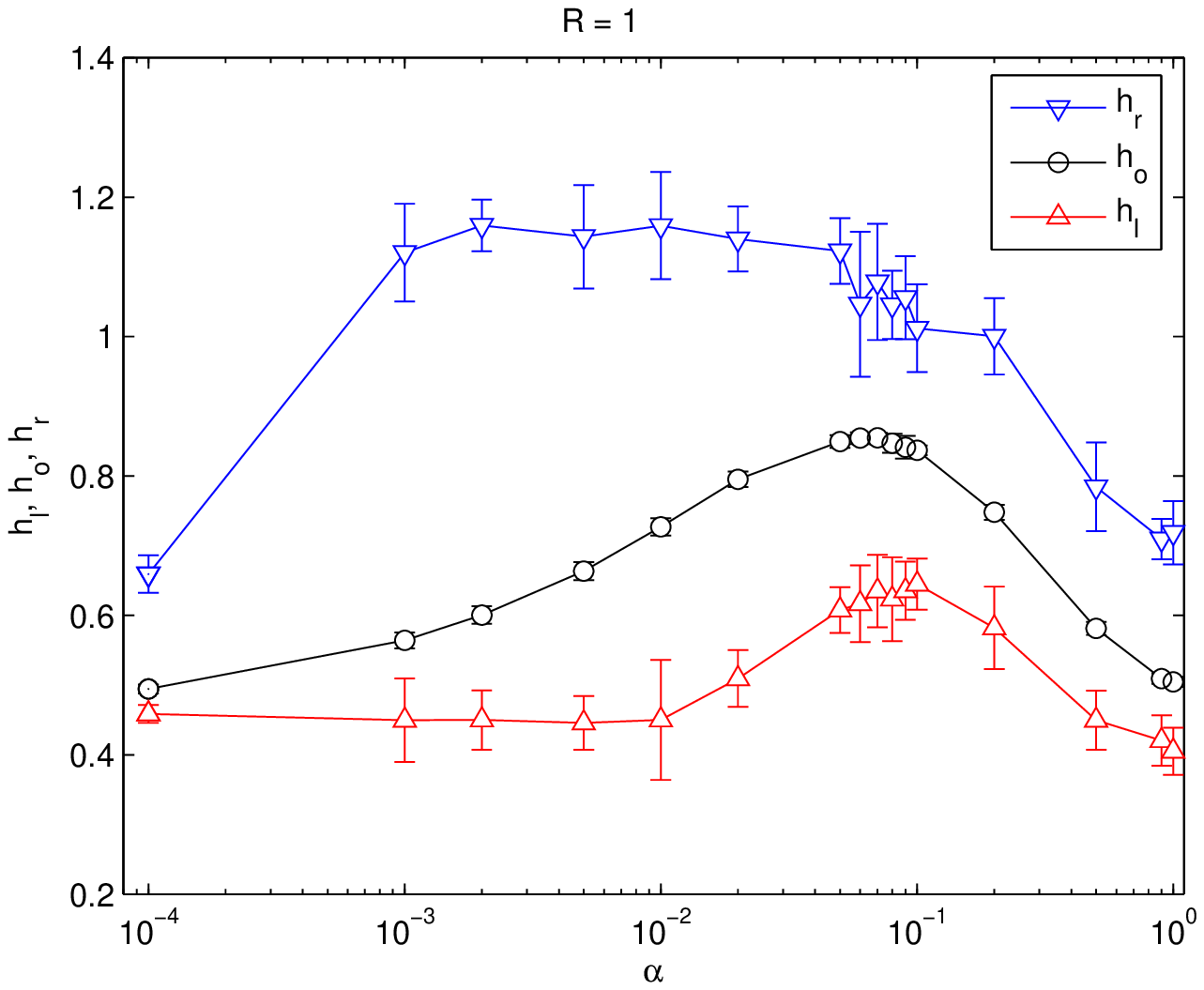} 
  \includegraphics[width=6.5cm]{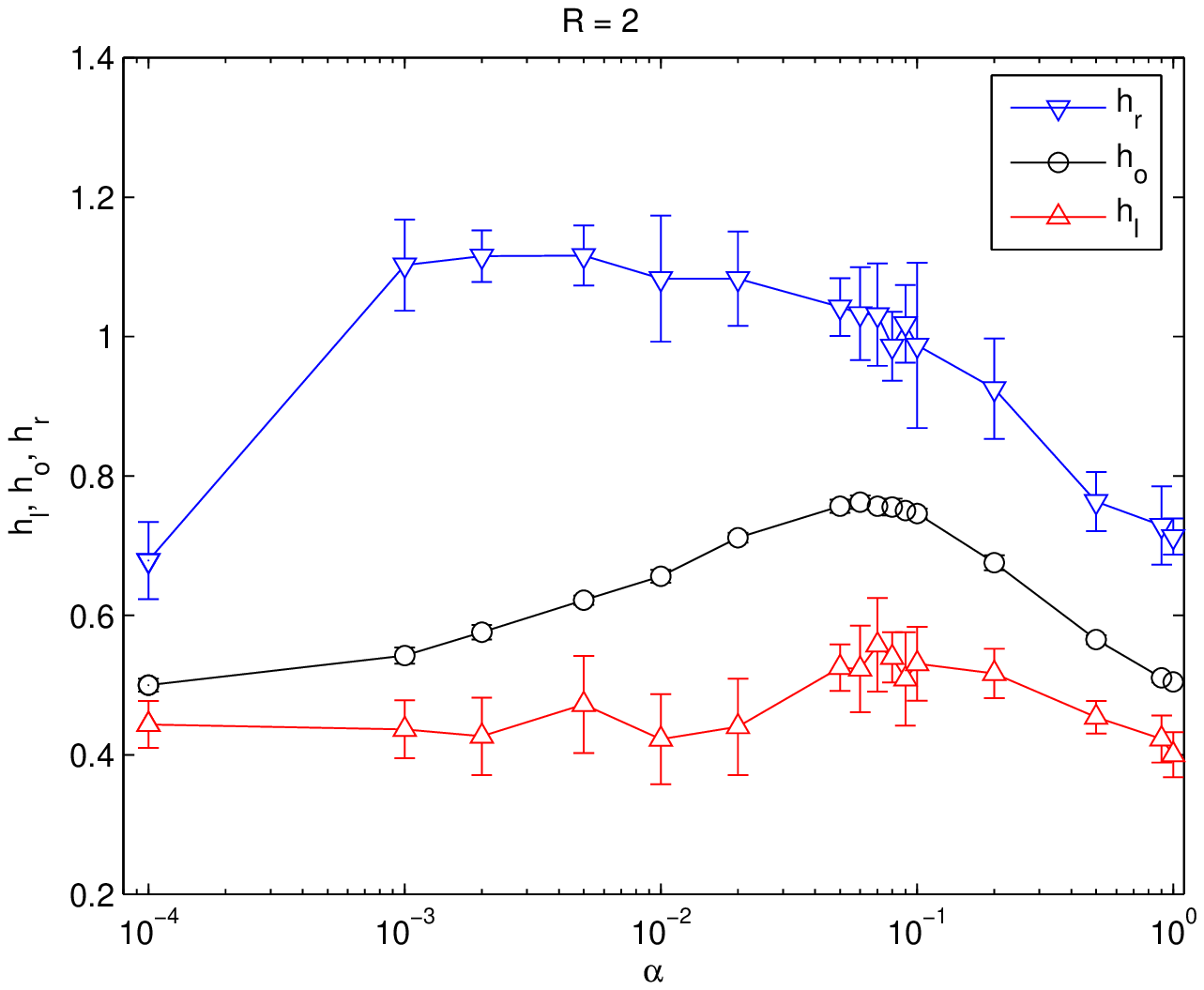} \\
  \includegraphics[width=6.5cm]{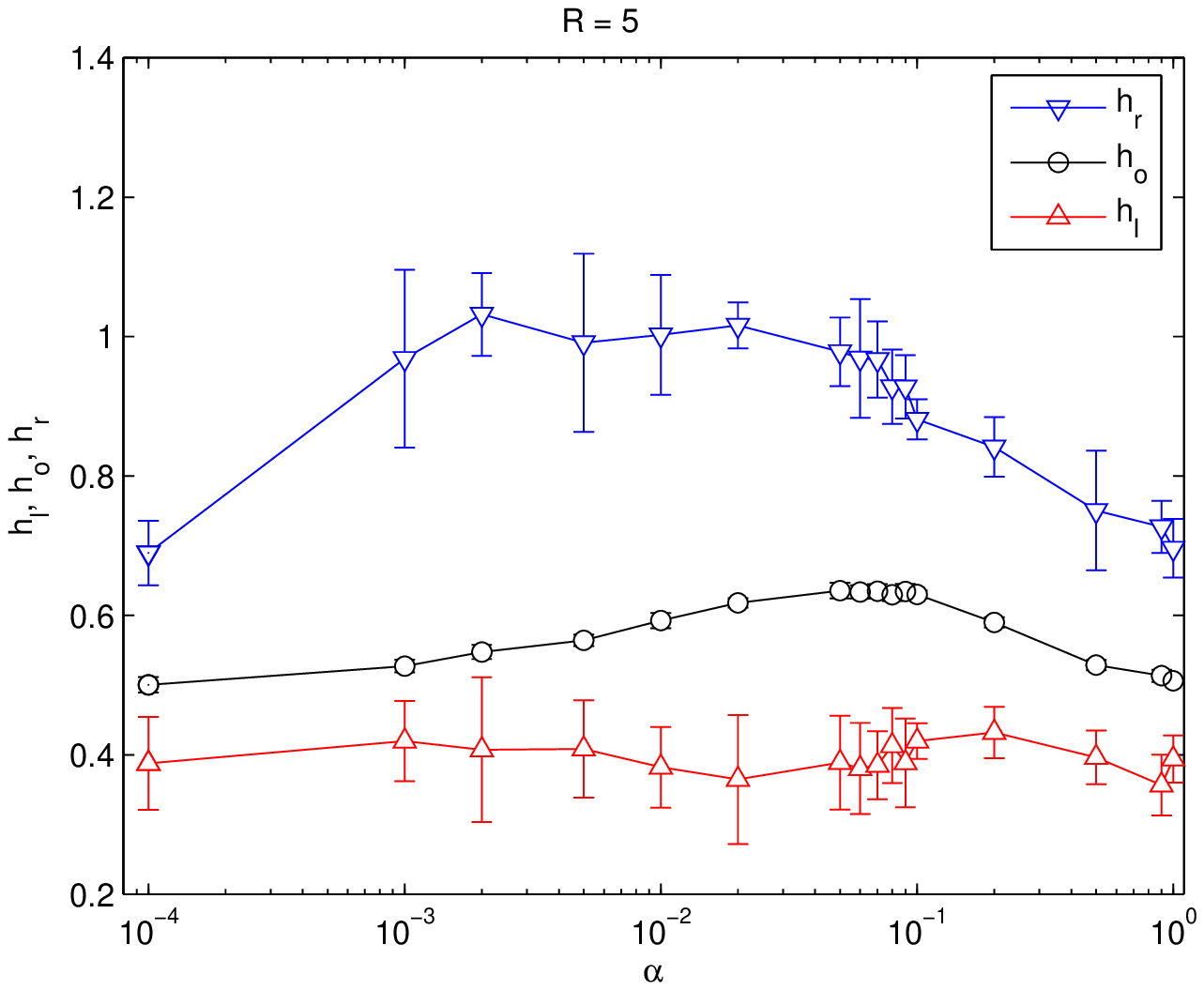} 
  \includegraphics[width=6.5cm]{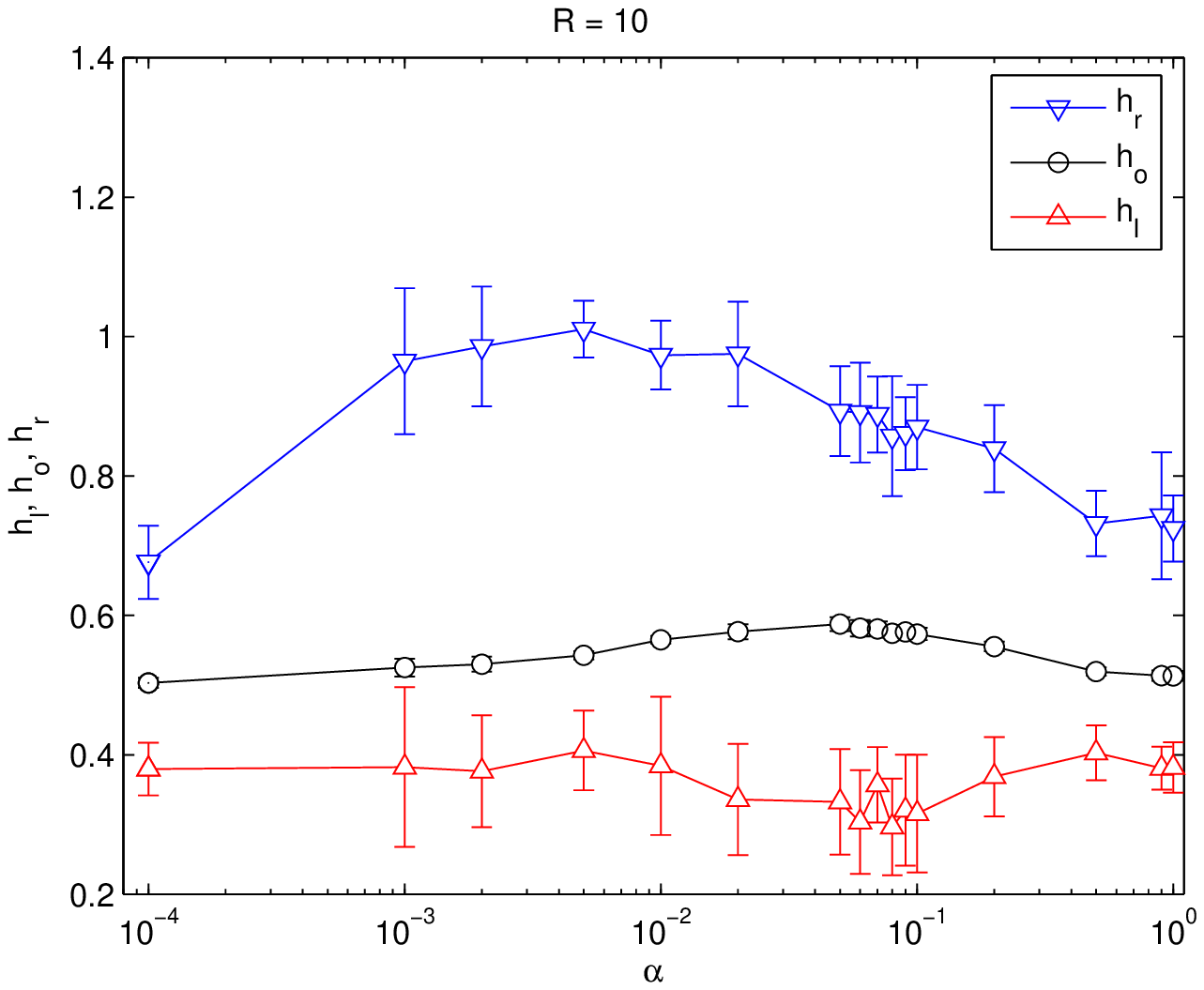}
  \caption{Multifractal spectrum dependence with $\alpha$ for (a) $R=1$, (b) $R=2$, 
(c) $R=5$ and (d) $R=10$.
The error bars are obtained from $10$ realizations.}
\label{pn}
\end{figure}

Now we will look for the multifractal properties of the model dynamics
as a function of the model parameters $R$ and $\alpha$.
The multifractal spectrum may be represented by its extrema points, i.e., 
its minima on the left, $h_l$, and on the right, $h_r$, as well as the maximum  (top), $h_0$.
It worths to note that wider the spectrum, i.e. bigger the difference between $h_r$ and $h_l$, 
more evident it is the multifractality character.
Fig. \ref{pn} shows that the 
multifractal spectrum is narrow for small and high values of $\alpha$.
The smallest value simulated was $\alpha = 0$, a system with no mean-reversal.
The highest, $\alpha=1$, is also non-realistic, since it represents a system
with no memory.
As it was already expected, these regions are clearly fractal, since 
the time series for the returns are probably close to gaussians (zero kurtosis).
For intermediary values, $\alpha\sim 0.01$, values close to what would be spected 
in real markets, the multifractal spectrum is wide, a sign of multifractality.
In this and all other figures we used time series of 32k points long --
for longer time series, the multifractal spectrum for small and high $\alpha \,$'s 
almost collapses into a single point, indicating a tendency to fractal behavior.

\begin{figure}[bbb]
 \includegraphics[width=12.5cm]{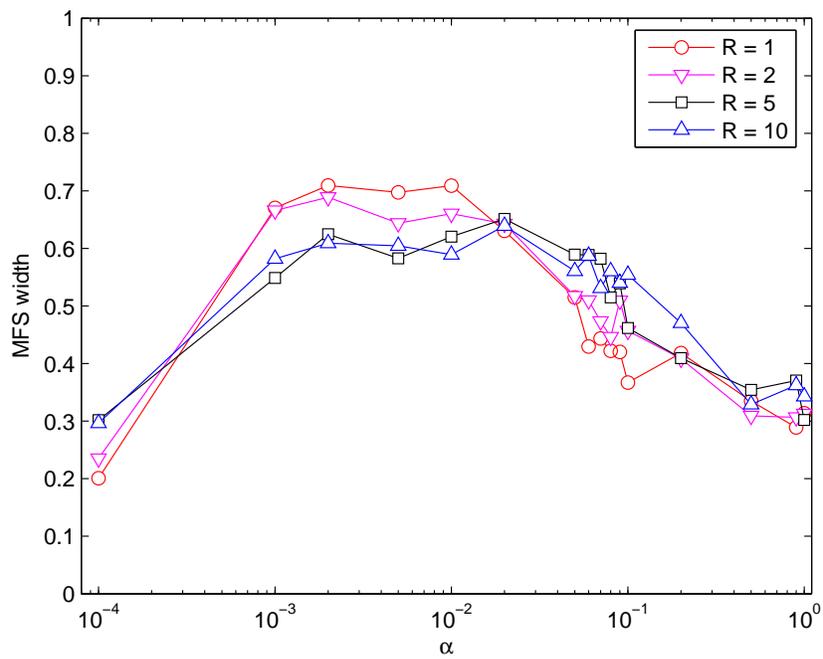}
  \caption{Multifractal spectrum width as a function of $\alpha$ for all four values of
$R$ studied. Surprisingly, all the four values of $R$ have the same width
dependence with $\alpha$, although their multifractal spectrum have different absolute values and
different values for $h(q=2)$, the Hurst exponent.}
\label{mfs_w}
\end{figure}

The multifractal character of the time series is more evident from the plot for
the spectrum width as a function of $\alpha$, shown in Fig. \ref{mfs_w}.
This figure shows the difference between the
maximum and the minimum values of $h$ for each value of $\alpha$, for all four 
values of $R$ studied ($R=1,\; 2,\; 5$ and $10$).
As $\alpha$ increases, the spectrum width goes from a fractal to a multifractal regime 
and then returns to a fractal one.
This transition is smooth exhibiting large fluctuations for $h_l$ and $h_r$ in the 
intermediary values of $\alpha$.
It worths to note that the point $h_l$ shows the scaling of the large fluctuations 
on the time series, while the $h_r$ captures the scaling of the small fluctuations.

\begin{figure}[bbb]
 \includegraphics[width=12.5cm]{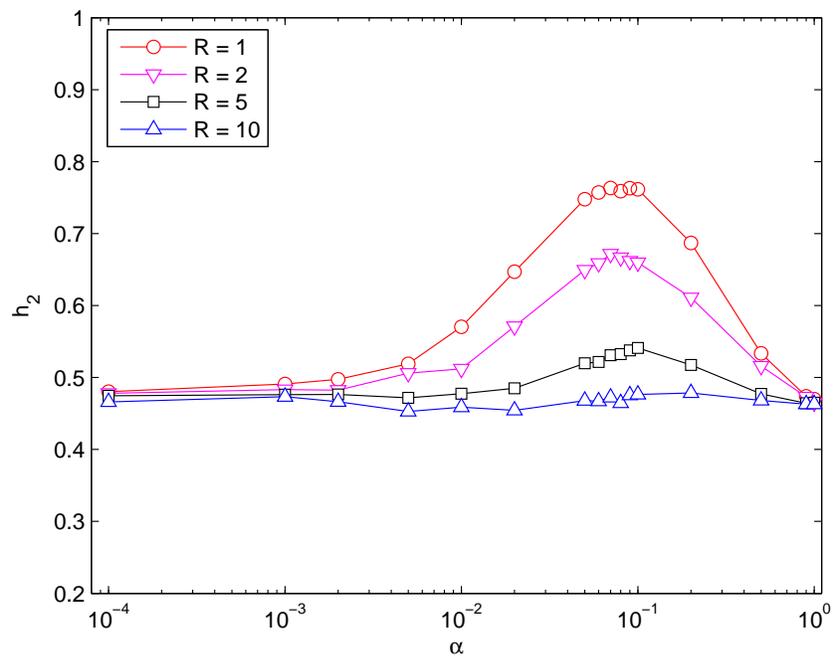}
  \caption{The Hurst exponent dependence with $\alpha$ for all the
values of $R$ simulated. Curves for increasing values of $R$ show smaller values of the Hurst
exponent.}
\label{hurst}
\end{figure}

Another representative point of the multifractal spectrum is the value of the Hurst
exponent, obtained from $h(q=2)$. 
As shown by the Fig. \ref{hurst}, the Hurst exponent for small and high values of $\alpha$
is close to $0.5$, the value expected for Brownian random walks.
For intermediate, more realistic values $\alpha \sim 0.01$, the time series 
becomes more persistent, as the Hurst exponent increases, coherent with the 
real persistent behavior of the market. 
We could still speculate about the interpretation of the meaning of the $\alpha$ parameter.
At intermediary values of $\alpha$, the parameters of the system have some mean-reversal, 
some memory and the series show increasing values of $H$, all features presented in the
real markets. Under this regime, a fluctuation that made the value of the parameters temporarily
larger will take longer to bring them back and, therefore, some the observation of 
persistent behavior makes sense. Under these circumstances, it makes more sense, to predict near
future behavior, to use smaller recent series.
This is not the case in the no memory and complete memory regimes.

\section{Conclusion}

The Random Paremeter model was build to explain the covariance structure of time series where 
non-stationarity might be an important feature, since it allows to implement non-stationarity in 
the parameters, while trivially respecting the properties of the covariance matrix. 
We have seen here that the model can also be used to 
explain other stylized facts, as the multifractal spectrum of financial series. By exploring the
PDFs of the generated time series, we have seen how to adapt the model to describe data of high and
low frequency; as $R$ is larger, the series match the behavior of high frequency series better. 
The multifractal spectrum appears when the mean-reversal term is not too large, something that is
compatible with real markets. This means that this model is a good choice in describing several
properties of the real series and, therefore, should be further investigated.

\label{final}

%\bibliography{multifractality}

\newpage

\end{document}